\documentstyle [12pt,a4p,epsfig,amsmath,multicol]{article}
\begin{document}
\vspace*{0.6cm}

\begin{center} 
{\normalsize\bf On some erroneous comments on the literature of neutrino oscillations
   in the website `Neutrino Unbound' of C.Giunti}
\end{center}
\vspace*{0.6cm}
\centerline{\footnotesize J.H.Field}
\baselineskip=13pt
\centerline{\footnotesize\it D\'{e}partement de Physique Nucl\'{e}aire et 
 Corpusculaire, Universit\'{e} de Gen\`{e}ve}
\baselineskip=12pt
\centerline{\footnotesize\it 24, quai Ernest-Ansermet CH-1211Gen\`{e}ve 4. }
\centerline{\footnotesize E-mail: john.field@cern.ch}
\baselineskip=13pt
 
\vspace*{0.9cm}
\abstract{A number of misleading or incorrect comments by C.Giunti
 on seven arXiv preprints that I have written on the theory of neutrino
   oscillations are discussed. The essential new features of my approach are
   also briefly reviewed}
\vspace*{0.9cm}
\normalsize\baselineskip=15pt
\setcounter{footnote}{0}
\renewcommand{\thefootnote}{\alph{footnote}}
\newline
PACS    03.65.Ta, 14.60.Pq, 13.20.Cz 
\newline
{\it Keywords ;} Quantum Mechanics,
Neutrino Oscillations.
\newline

\vspace*{0.4cm}

   In providing and maintaining the website `Neutrino Unbound' ~\cite{NU} C.Giuni is
 providing an enormous service to the neutrino physics community. 
  In the comments on the literature in the `Theory of Neutrino Oscillations' section,
  however, there is a mixture of personal scientific judgements on a limited sub-set
  of subjectively `important' issues as well as a systematic omission of any comment
  on certain key contradictory
  arguments presented in the literature. There are also factual errors in the comments.
 In an e-mail exchange
  I had with Giunti on the subject of his webpage he claimed that, in any case, only his
  personal opinion was expressed. However language is sufficiently rich to discriminate clearly
  between a fact that one knows (e.g. that such-and-such a subject was discussed in 
  such-and-such a paper leading to such-and-such a conclusion) and one's personal
   judgement of the correctness, or not, of some
  scientific theory. In the following I try my best to make this discrimination.
  \par In the papers I have written recently on the subject of the basic theory
   of neutrino oscillations ~\cite{JHF1,JHF2,JHF3,JHF4,JHF5} I make different initial
   physical assumptions (the most important one of which I rediscuss at the end of this note)
   to those made by essentially all authors on the subject from Pontecorvo onwards. It is then
   not surprising 
  that I obtain different results. These different basic assumptions that underlie my calculations
   are not pointed out in any of Giunti's comments on my papers. Instead it is wrongly claimed
   in several different places that there are `mistakes' in my calculations. Giunti also claims
   in his comment on one of his own papers that the correctness of the standard formula for the
   oscillation phase has been `shown', when, indeed, it has not. Only different initial assumptions
   are made to those in my papers and different results are obtained.
   \par I have shown in a recent paper~\cite{JHF4}
  that a clear and simple experimental test exists, comparing oscillations of neutrinos originating
  either from pion or kaon decays, to discriminate unambigously between the `standard' formula
  for the oscillation phase~\cite{KaysPDG} and my predictions.
  \par In the following, I copy {\it verbatim} some of the comments on the literature from 
   `Neutrino Unbound' and add my remarks to them.
   \newline
   \par{ \bf The description of neutrino and muon oscillations by interfering amplitudes of
    classical space-time paths, Field, J. H.,
     hep-ph/0110064}.
    { \sf Comment: Again the factor of 2 mistake in the phase!
      This error has been explained in [Carlo Giunti, Chung W. Kim, Found. Phys. Lett. 14 (2001) 213-229,
     hep-ph/0011074 ]. [C.G.].}
    \newline 
     \par The factor 2 difference in the contribution of neutrino propagation to the oscillation
   phase results from the constraint of space-time geometry on the neutrino trajectories. 
    Since the different mass eigenstates are detected at a common space-time point, travel the
   same distance and have different velocities, they are necessarily produced at different times
   in the interfering amplitudes. This assumption was not discussed in the paper of Giunti and Kim
   cited above. The statement `This error has been explained in...' is therefore false. 
    In fact, it is simply assumed in the paper of Giunti and Kim (in Eqn(22)) that the
     neutrinos have equal velocities
    in space-time
     (i.e. a common production time) but different kinematical velocities $p/E$. As explained in
     much detail in Reference~\cite{JHF5} these manifestly incompatible assumptions
     (equal space-time velocities but different kinematical velocities) underestimate by a
   factor of 2 the contribution of neutrino propagation to the oscillation phase and 
    result in the neglect of an important contribution to the oscillation phase from the
   propagator of the source particle. It is also shown in Reference~\cite{JHF5} that the
   same initial assumptions, when made in the standard Gaussian wave-packet treatment, leads to the 
   same underestimation of the oscillation phase.
  \newline
   \par{ \bf A covariant path amplitude description of flavour oscillations: The Gribov-Pontecorvo phase
   for neutrino vacuum
     propagation is right, Field, J. H., hep-ph/0211199.}
  {\sf Comment: Same as [Field, J. H.,
     hep-ph/0110064] and [Field, J. H., hep-ph/0110066].
     Mistake explained in [Carlo Giunti, Chung W. Kim, Found. Phys. Lett. 14 (2001) 213-229,
     hep-ph/0011074] and Giunti, C., Physica Scripta 67 (2003) 29-33, hep-ph/0202063]. 
     The attribution of the "factor of two" mistake to Gribov and Pontecorvo is an historical
    aberration. 
      The claim that the "factor of two" discrepancy in the Gribov and Pontecorvo paper
 [Gribov, V. N., Pontecorvo, B., Phys. Lett. B28 (1969) 493.] 
     was unnoticed before
     [ Field, J. H., hep-ph/0211199] is pure fantasy. The fact is that nobody speculated about it. 
     [C.G.].} 
     \newline
     \par There is no `mistake', unless the assumption that the space-time and kinematical
      velocities are the same is one. In any case, nothing is `explained' in the references
     cited. The key assumption, that is the basis of my calculations, is not even
     discussed in the first of them. In the second one, it is discussed, and it is claimed 
     that an analagous analysis of the optical Young double slit experiment excludes the possiblity
     of different production times. However the path amplitude analysis presented is 
      incorrect as it does not take into account the contribution to the interference
       phase of the propagator of the source. All this is explained in Reference~\cite{JHF5}.

      \par Maybe lots of people knew about
     the factor 2 difference in Gribov and Pontecorvo, but, as I am not a clairvoyant, there
    is no way I could know this. All I said was that, as far as I know, the difference
   was never discussed in the literature.
    \newline
    \par { \bf The phase of neutrino oscillations,
     Giunti, C., Physica Scripta 67 (2003) 29-33, hep-ph/0202063.}
    {\sf  Comment: It is shown that the standard phase of neutrino oscillations is correct,
     refuting the claims of a factor of two
     correction presented in [ S. De Leo, G. Ducati, P. Rotelli, Mod. Phys. Lett. A15 (2000)
     2057-2068, hep-ph/9906460,   Field, J. H.,
     hep-ph/0110064., Field, J. H., hep-ph/0110066.]. 
     The wave packet treatment of neutrino oscillations presented in [C. Giunti, C. W. Kim, U. W. Lee,
     Phys. Rev. D44 (1991) 3635-3640, C. Giunti, C. W. Kim, Phys. Rev. D58 (1998) 017301,
     hep-ph/9711363.]
     is improved taking into account
     explicitly the finite coherence time of the detection process.}
      \newline   
      \par The argument given in this paper to refute the results obtained in my papers, as cited above,
      is based on an incorrect path-amplitude analysis of the Young double slit experiment 
      in optics. This analysis is discussed in  Reference~\cite{JHF5} . In Giunti's comment on this 
      paper (see above) there is no mention that his analysis has been shown in it to be incorrect,
       and that the correct one gives the same result as the classical wave theory of light. In any case,
     it is not possible `That it is shown that the standard phase of neutrino oscillations is correct'.
     One can only discuss the correctness, or not, of the mathematical derivation given some initial
     set of physical assumptions. To prove the correctness of the result, that of the initial
     asumptions must also be demonstrated. In all papers, authored or co-authored by Giunti, this
     initial assumption is that the different mass eigenstates have different kinematical velocities
     but the same space-time velocity. This assumption has never been proved. My personal
     opinion is that the assumption is incorrect, but I cannot prove this. The experiment I
     suggested in Reference~\cite{JHF4} can, however, resolve, unambigously, this question.
     \newline
     \par {\bf Coherence in Neutrino Interactions, C. Giunti, hep-ph/0302045}.
      { \sf From the abstract: The claim in [Field, J. H.,
     hep-ph/0301231.] is refuted in a pedagogical way.}
      \newline 
      \par As shown in Reference~\cite{JHF1} the modification proposed in this paper to the
       Standard Model amplitude for pion decay avoids the experimental constraint
       of Reference~\cite{JHF4} demonstrating the incoherence of the production of
       different mass eigenstates in 
       pion decay, but also
        forbids the possibility of 
        neutrino oscillations following pion decay!
        The derivation of the standard 
        oscillation phase requires, instead, coherent production of the
        different mass eigenstates. Since neutrino oscillations following pion
         decay apparently exist, the predictions of this 
         paper are experimentally excluded, so that it cannot, in any way, `refute' the results
        of Reference~\cite{JHF4}. Giunti's only comment on Reference~\cite{JHF1}
        is a quotation from the Bible, the
        meaning and relevance of which is unclear to me.
    \newline
     \par The crucial assumption that distinguishes my treatment of the quantum mechanics
      of neutrino oscillations from all of the papers cited in the section `Theory of Neutrino
      Oscillations' of `Neutrino Unbound' (with the exception of Reference~\cite{DLDR}) is
      that mentioned above in connection with Giunti's comments on my paper hep-ph/0110064.
      (now superseded by hep-ph/0303151). This is the condition that the velocities of the
       neutrino mass eigenstates are described consistently both kinematically,
       and in space-time. That is to say that the neutrinos propagate over macroscopic
     distances as classical particles. Such an assumption is fundamental to Feynman's
     path amplitude formulation of quantum mechanics. To each such classical trajectory
    corresponds a probability amplitude. That the Feynman path amplitude description
     gives a correct quantitative description of nature has been demonstrated in many
     atomic physics experiments involving space-time interference effects, some of which
       are discussed in Reference~\cite{JHF3}.
     \par The universal assumption that is made when the `standard' oscillation 
      phase is derived is that the different neutrino eigenstates are produced at
      a unique time since they are assumed to be different components of a coherent
       `lepton flavour eigenstate'. As first pointed out by Shrock~\cite{Shrock1}, in the
        Standard Model the different mass eigenstates are actually 
       produced {\it incoherently} in different physical processes, and so not necessarily at the
        same time, which is the case when a coherent state is produced. It is shown in
       Reference~\cite{JHF4} that the production of such a coherent state in pion decay is
       excluded by the measured $e/\mu$ branching ratio.
      \par The other important point where my approach differs from many of the other 
      treatments in the literature is in the importance accorded to the introduction
     of {\it ad hoc} Gaussian `wave-packets' in the description of neutrino oscillations.
      An in-depth criticicism of such approaches is presented in Reference~\cite{JHF5}.
      In a later paper on the same subject by Giunti~\cite{Giunti1} not only 
      were no counter arguments to my criticisms given, but my paper was not even cited!
      As discussed in Reference~\cite{JHF3} there are indeed momentum
      wave packets in the exact path amplitude description of pion decay due to the
     unobserved smearing of the physical mass of the recoiling muon. The effect on the 
     formula for the oscillation phase is, however, completely negligible. 
     \par Another subject that has caused much ink to flow and contradictory positions
      to be adopted is the role of different assumptions concerning the calculation
      of the kinematical velocities of the neutrinos: exact energy-momentum
      conservation, equal momenta or equal energies. It is pointed out in Reference~\cite{JHF5}
      that the same result, at O($m^2$), is obtained whatever assumption is made, so that
     unless O($m^4$) corrections to the oscillation phase are of interest, this debate is 
     an irrelevant one.

\end{document}